\documentclass[fleqn,usenatbib]{mnras}

\usepackage[T1]{fontenc}

\DeclareRobustCommand{\VAN}[3]{#2}
\let\VANthebibliography\thebibliography
\def\thebibliography{\DeclareRobustCommand{\VAN}[3]{##3}\VANthebibliography}

\usepackage{graphicx}	
\usepackage{amsmath}	
\usepackage{amssymb}	

\newcommand{\Msunpc}{\,M$_\odot$\,pc$^{-1}$} 
\newcommand{\kms}{\,km\,s$^{-1}$} 
\newcommand{\K}{\,K} 
\newcommand{\gccm}{\,g\,cm$^{-3}$} 
\newcommand{\Myr}{\,Myr} 
\newcommand{\pc}{\,pc} 

\title[Filament collapse]{Filament collapse: a two phase process}

\author[E. Hoemann et al.]{
Elena Hoemann,$^{1,2}$\thanks{E-mail: hoemann@usm.lmu.de}
Stefan Heigl$^{1,3}$
and Andreas Burkert$^{1,2,3}$
\\
$^{1}$Universitäts-Sternwarte, Ludwig-Maximilians-Universität München, Scheinerstr. 1, 81679 Munich, Germany \\
$^{2}$Max-Planck Institute for Extraterrestrial Physics, Giessenbacherstr. 1, 85748 Garching, Germany \\
$^{3}$Excellence Cluster ORIGINS, Boltzmannstrasse 2, 85748 Garching, Germany
}

\date{Accepted 2023 March 16. Received 2023 March 16; in original form 2022 March 14}

\pubyear{2023}

\usepackage{newtxtext,newtxmath}
\begin{document}
\label{firstpage}
\pagerange{\pageref{firstpage}--\pageref{lastpage}}
\maketitle

\begin{abstract}
  Using numerical simulations, we investigate the gravitational evolution of filamentary molecular cloud structures and their condensation into dense protostellar cores. One possible process is the so called `edge effect', the pile-up of matter at the end of the filament due to self-gravity. This effect is predicted by theory but only rarely observed. To get a better understanding of the underlying processes we used a simple analytic approach to describe the collapse and the corresponding collapse time. We identify a model of two distinct phases: The first phase is free fall dominated, due to the self-gravity of the filament. In the second phase, after the turning point, the collapse is balanced by the ram pressure, produced by the inside material of the filament, which leads to a constant collapse velocity. This approach reproduces the established collapse time of uniform density filaments and agrees well with our hydrodynamic simulations. In addition, we investigate the influence of different radial density profiles on the collapse. We find that the deviations compared to the uniform filament are less than 10\%. Therefore, the analytic collapse model of the uniform density filament is an excellent general approach.
\end{abstract}

\begin{keywords}
stars:formation -- ISM:kinematics and dynamics -- ISM:structure
\end{keywords}



\section{Introduction}
\label{sec:introduction}

  Observations show that the interstellar medium (ISM) is dominated by filamentary structure. Filaments are the cylindrical and cold density enhancements of the molecular ISM. They have been found in very different environments from large scales (more than 100\pc) like infrared dark clouds \citep{Goodman2014, Mattern2018} down to small, (sub-) parsec scales \citep{Molinari2010, Andre2010, Hacar2013, Schmiedeke2021} at the current detection limit. It is well established that filaments in the ISM are the places where star-formation happens \citep{Schneider1979}, since most of the prestellar cores are found within these dense, cold environment \citep{Arzoumanian2011, Koenyves2015}. However, there remain many open questions about their creation, evolution and fragmentation. 
  
  A particular example is the so called `edge effect'\citep{Bastien1983}: The acceleration along a finite filament due to its self-gravity has a strong increase at its edges because of its elongated structure \citep{Burkert2004, Hartmann2007, Li2016}. Thus, the collapse leads to a pile-up of matter in the end regions. This effect has been studied theoretically \citep{Rouleau1990, Arcoragi1991, Bastien1991, Hanawa1994} and has been detected in several observational studies \citep[e.g.][]{Zernickel2013, Kainulainen2016, Dewangan_2019, Bhadari_2020, Yuan2020, Cheng2021}. Since there is no hydrostatic solution for the end of a long filament which could prevent the edge effect, it is expected to occur quite often. In contrast, \citet{Seifried2015} showed that filaments with an initial inner density enhancement collapse centrally. A similar setup was investigated by \citet{Keto2014} for short filaments with axis ratio of 3:1. They find that the resulting core shows complex pattern of sonic oscillations. Although, such density distributions are not observed in low line-mass filaments \citep{Roy2015}, the oscillations in cores have been detected \citep{Redman2006, Aguti2007}. In addition, there has been a recent study by \citet{Heigl2021} where the edge effect is suppressed in a filament forming in a colliding flow, however, not every filament shows signs of accretion. Thus, the edge effect is expected to happen much more often than it is detected.
  
  Understanding the edge effect is crucial for setting a limit for filament lifetimes because the collapse timescale is given by the time on which the two end regions collapse into a single point. Collapse times have been calculated by \citet{Toala2011, Pon_2012} and \citet{Clarke2015}. However, they all rely on the acceleration of uniform density filaments as calculated by \citet{Burkert2004}, a density profile which is neither observed nor theoretically expected.
  
  The aim of this study is to get a more accurate impression of the longitudinal collapse of a filament and its corresponding collapse timescale. Therefore, we present an analytical model to calculate the evolution of the collapse and to explain the collapse timescale found in \citet{Clarke2015}. We used a simple approach consisting of two phases: an accelerated one, dominated by gravitational free fall, which turns into a collapse with constant velocity at the point where ram pressure sets in. Additionally, the influences of the filament's radial profile are analyzed, which has only minor contributions. Thus, the two phase model of collapse and force equilibrium for a uniform density filament can be used in general to determine collapse timescales of observed filaments.  
  
  The paper is organized as follows: After introducing the basic principles of filaments in hydrostatic equilibrium (Section \ref{sec:basic_priniciples}), the theoretical background of  the edge effect is introduced (Section \ref{sec:edge_effect}). The simulation is discussed in Section \ref{sec:simulating}. In Section \ref{sec:collapse_uniform_density} the two phase collapse model of a uniform density filament is derived and in Section \ref{sec:adjustments} we show a comparison of different profiles and their influence on the collapse time. The results are discussed in Section \ref{sec:discussion} and conclusions are drawn in Section \ref{sec:conclusion}.

\section{Filaments in hydrostatic equilibrium}
\label{sec:basic_priniciples}

  Following \citet{Stodolkiewicz1963} and \citet{Ostriker1964} a filament in hydrostatic equilibrium has the following radial density distribution:
  \begin{align} \label{eq: Ostriker}
      \rho(r) = \rho_c \left[ 1 + \left( \frac{r}{H} \right)^2 \right]^{-2}
  \end{align}
  with $\rho_c$ being the central density. $H$ is the scale height, given by
  \begin{align}
      H^2 = \frac{2c_\mathrm{s}^2}{\uppi G \rho_c},
  \end{align}
  $G$ being the gravitational constant and $c_\mathrm{s}$ the sound speed \citep[0.19\kms\ for a temperature of 10\K\ and a mean molecular weight of 2.36, compare][]{Fischera2012}. Filaments in the ISM do not extend until infinity since they are constrained by an external pressure $P_{\mathrm{ext}}$. This cuts off the filament at pressure equilibrium between the boundary pressure $P_\mathrm{b}$ and the external pressure $P_\mathrm{b}=P_{\mathrm{ext}}$ at density $\rho_\mathrm{b}$. Its line-mass is given by
  \begin{align}
      \mu = \frac{M}{l}
  \end{align}
  the mass $M$ divided by the filaments length $l$. Integrating the hydrostatic profile until infinity gives the maximal line-mass for which a hydrostatic solution of the density profile exists and above which filaments collapse radially \citep{Fischera2012}. The critical value for $T=10$\K\ is
  \begin{align}
      \mu_{\mathrm{crit}} = \frac{2c_\mathrm{s}^2}{G} \approx 16.4 \text{\Msunpc}.
  \end{align}
  The criticality is then given by the fraction of the actual line-mass to critical line-mass:
  \begin{align}
      f = \frac{\mu}{\mu_{\mathrm{crit}}}.
  \end{align}
  Thus, a filament is considered supercritical when it exceeds $f=1$ where no hydrodynamic solution can be found and the filament would start to collapse radially. Together with the central density the cirticality determines the boundary density
  \begin{align}
      \rho_\mathrm{b} &= \rho_\mathrm{c} (1-f)^2
  \end{align}
  and the corresponding filament radius $R$
  \begin{align}
      R &= H \left( \frac{f}{1-f} \right)^{1/2}.
  \end{align}

\section{Edge effect}
\label{sec:edge_effect}

\begin{figure}
      \includegraphics[width=0.48\textwidth]{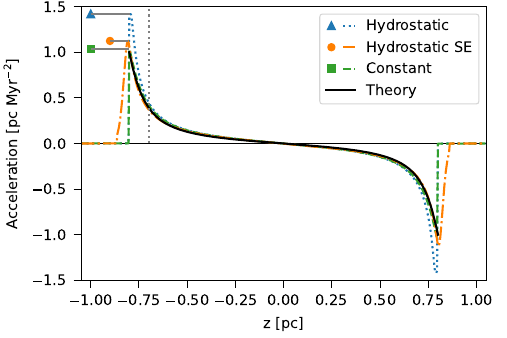}
      \caption{Acceleration along a filament with $A=8$,   $R=0.10$\pc, $f=0.7$ at its radial centre $r=0$. The black line shows the analytical solution of a uniform density filament given by Equation \ref{eq:accelerationCollapseFilament} \citep{Burkert2004}. The coloured lines show the numerical determined accelerations for various density profiles: The constant density filament is given in dashed-green, the hydrostatic filament in dotted-blue, and the hydrostatic filament with soft edges hydrostatic SE (Equation \ref{eq:soft_edge}) in dashed-dotted orange. At the end of the filament the acceleration is profile dependent and thus, deviates from the theoretical prediction. However, inside the filament and especially at the approximated center of mass of the end region (marked by the grey dotted line) the theoretical approximation fits well for all profiles.}
      \label{fig:acceleration}
  \end{figure}

  Although there exists a hydrostatic solution for the radial profile of a filament (as discussed in Section \ref{sec:basic_priniciples}), for the z-direction there are only solutions for elongated cores \citep{Lizano1989, Tomisaka1991, Cai2010}. This means there is no hydrostatic solution for the main axis of a filament with a considerable large aspect ratio $A=l/(2R)$. Thus, there is no possible profile for which the filament's end is stable. Therefore, it is expected that every isolated filament with finite length should collapse along its axis. The acceleration along the z-axis of a cylindrical filament with uniform density distribution (at r=0) was already investigated by \citet{Burkert2004}
  \begin{align} \label{eq:accelerationCollapseFilament}
	a = -2 \pi G \bar{\rho} \left[ 2z - \sqrt{\left( \frac{l}{2} + z \right)^2+R^2} + \sqrt{\left( \frac{l}{2} - z \right)^2+R^2} \right],
  \end{align}
   with $\bar{\rho}$ being the uniform density and $z$ being the position along the filament. An example of the acceleration is given by the black solid line in Figure \ref{fig:acceleration}. The steep increase in acceleration leads to a pile-up of matter in clumps forming at the ends of the filament. These move inward and finally merge in the center. This end dominated collapse is what is called the `edge effect' \citep{Bastien1983}. Of course the two end points can collapse gravitationally when they reach their local Jeans mass. In this case however, the protostellar cores will continue to move inwards until stellar feedback will become active. We do not model the effects of stellar feedback as it is beyond the scope of this paper and concentrate on the primary effect of gravitational collapse. Here we focus on the motion along the z-axis neglecting the compression by self-gravity in the x-y plane. We however explore different density structures in the x-y plane and demonstrate that the evolution in the z-direction is to first approximation independent of this detail. The acceleration at the exact end of the filament for $L\gg R$ can be approximated by: 
  \begin{align} \label{eq:accerelation_end}
      a = \pm 2\uppi G \bar{\rho} R.
  \end{align}
  This shows that the acceleration at the end of a filament is independent of its length.
  
  In Figure \ref{fig:acceleration} the acceleration for different profiles, measured in the simulations, is also depicted in comparison to the theoretical expectation. The acceleration of a constant density filament is shown in green, which matches the theoretical expected values. The hydrostatic filament, given by the blue dotted line, experiences a similar acceleration inside the filament. However, the maximum acceleration at the exact end of the filament is larger, as indicated by the grey horizontal line, marked with the blue triangle. The filament has a sharp cut at the end to match the cut-off of the uniform density filament. However, more soft edges (orange), defined as 
  \begin{align} \label{eq:soft_edge}
    \rho _{\mathrm{end}} (r,z) = \rho (r) \mathrm{sech} \left( 2 \frac{|z|-l/2}{H} \right)^2
  \end{align}
  for $|z|>l/2$, had only minor influences, besides lowering the acceleration at the exact end. The grey dotted line is situated one radius away from the end inside the filament, which was used in \citet{Hoemann2021} as centre of mass of the end region. At this position the variations are negligible for all three cases. 
  
  As already mentioned, there is no hydrostatic solution for a cylinder with a finite length and considerably large aspect ratio, thus it is theoretically expected that every cylindrical filament should experience the edge effect. Although there are some observations of the edge effect \citep{Zernickel2013, Kainulainen2016, Dewangan_2019, Bhadari_2020, Yuan2020, Cheng2021}, in most cases no overdensities are detected at the end of filaments. This leads to a big puzzle: What suppresses the edge effect? A possible scenario is presented in \citet{Heigl2021} where the filament is formed by a constant radial accretion and the end regions are continuously fed with new material. Therefore, no real edge effect occurs. Nonetheless, an accretion flow is only detected seldom. It also has been shown that density fluctuations in the centre can lead to a central collapse mode \citep{Seifried2015}. However, the fluctuations need to be rather large, about a factor of three, which is normally not detected \citep{Roy2015}. Whether the edge effect is likely to be observed depends on its timescale. As in the case of filaments with small axis ratios, the collapse happens so fast that one hardly observes this phase. The complex, long-wavelength sonic oscillations of protostellar cores could then be the only detectable signature of this collapse phase \citep{Keto2014}. 
  This and the question whether one can prevent the edge effect requires a more detailed understanding of filament collapse which is the aim of this study.
  
  We already demonstrated in an earlier work \citep{Hoemann2021} that the timescale of the edge effect is dominated by the filament's criticality and its central density. This is the time needed to form self-gravitating cores, defined as a local over-density with $f>1$, at the edge of a filament:
  \begin{align} \label{eq:edge}
      t_{\mathrm{edge}} = \sqrt{ \frac{1.69\times 10^{-20}\mathrm{g\,cm^{-3}}}{f\rho _\mathrm{c} }} \mathrm{\Myr}.
  \end{align}
  In order to describe the overall collapse time, i.e. the time when the two end regions merge, several different approaches have been made: \citet{Toala2011, Pon_2012, Clarke2015}. The most recent one by \citet{Clarke2015} found a description of the collapse time for long filaments fitting SPH simulations:
  \begin{align} \label{eq:clarke}
      t_{\mathrm{col}} = \frac{0.49+0.26A}{\sqrt{G\bar{\rho}}}
  \end{align}
  where $A$ represents the aspect ratio $A \equiv l/(2R) \gtrsim 2$. They also found that the core reaches a terminal velocity after about 1\Myr\ of acceleration due to ram pressure. The detected terminal velocities showed nearly no dependence on the aspect ratio.
  
  However, a detailed analytic derivation of the collapse time (Equation \ref{eq:clarke}) has not yet been achieved.

\section{Simulating the filament collapse}
\label{sec:simulating}

  In order to validate our analytic study, we performed hydrodynamic simulations with the adaptive-mesh-refinement code RAMSES \citep{Teyssier_2002}. The Euler equations are solved in their conservative form by using a second-order Gudonov solver. We utilized the MUSCL \citep[Monotonic Upstream-Centered Scheme for Conservation Laws,][]{Leer1979}, the HLLC-Solver \citep[Harten-Lax-van Leer-Contact,][]{Toro1994} and the MC slope limiter \citep[monotonized central-difference,][]{Leer1979}.
  
  We simulated filaments with different density profiles, characterized by their aspect ratio $A$ (length divided by twice the radius), their radius $R$ and their criticality $f$. We used a grid refinement between level 7 (128$^3$ cells) and level 9 (512$^3$ cells) with open boundary conditions. Since the boxes were chosen to be 0.5\pc\ bigger than the inserted filament, we had a resolution outside the filament (low refinement) of $3.90\times 10^{-2}$\pc\,-\,$1.02\times 10^{-2}$\pc\ and inside (high refinement) from $9.8\times 10^{-3}$\pc\ to $2.5\times 10^{-3}$\pc. The outside density was set to $\rho_{\mathrm{ext}}=3.92\times 10^{-23}$\gccm (equivalent to 10 particles per cm$^3$) in pressure equilibrium to the boundary of the filament, to have no influence of accretion effects. Because we are only interested in the longitudinal collapse and want to suppress radial contributions, radial velocities were set to zero (see Section \ref{sec:collapse_uniform_density} for further discussion).
  
\begin{figure}
      \includegraphics[width=0.48\textwidth]{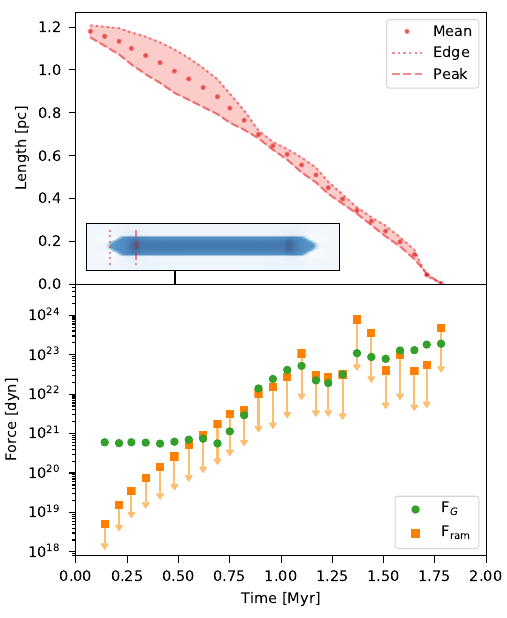}
      \caption{Length and force evolution of a hydrostatic filament with $A=12$, $R=0.05$\pc, $f=0.5$. \\
      Upper panel: The dotted red line depicts the evolution of the edge (where the line-mass decreases by a factor of three). The dashed red line follows the density peak. The mean is given by the red dots. The two positions are depicted by the small snapshot of the simulation in the left lower corner at $t=0.48$\Myr. In the following plots the outer lines (dotted, dashed) are only represented by the shaded region. \\
      Lower panel: The gravitational force, that the core experiences, is shown by the green dots, and the ram pressure acting against the collapse, is depicted as orange squares. First, the collapse is dominated by the gravitational force, which is counterbalanced by the ram pressure later on. The determination of the ram pressure is difficult in that case. Thus, the marker depicts the highest ram pressure value and the end of the arrow the mean value in the direct surrounding of the filament.}
      \label{fig:length_force}
  \end{figure}

  An example simulation is shown in Figure \ref{fig:length_force} (upper panel) with a radius of $R=0.05$\pc, a criticality of $f=0.5$ and an aspect ratio of $A=12$. The time evolution of the length can be traced by two values: distance between two peaks in the line-mass distribution (location of the end clump, `Peak', dashed line) or the actual end of the filament where the line-mass drops below a third of the initial value (`Edge', dotted line), as indicated in the subplot of the filament at $t=0.48$\Myr. The mean is given by the red dots, whereas the coloured area depicts the range between `Edge' and `Peak'. Surprisingly, the position of the core (dashed line in the upper panel of Figure \ref{fig:length_force}) follows a rather linear trend which is caught up by the outer end (small dotted line) over time. This indicates that there are two distinct phases in the collapse of a filament. The first part can be interpreted as a free fall as explored in \citet{Hoemann2021} which defines the epoch of formation of edge cores, namely that the edge cores become supercritical. At some point the acceleration is counterbalanced by the ram pressure experienced by the edge and a uniform movement sets in, which was already observed by \citet{Clarke2015} in their Figure 4a. This is also the trend seen in our simulation, see Figure \ref{fig:length_force} lower panel. There, green dots represent the gravitational force and the orange squares show the force due to ram pressure. The gravitational force is calculated by the gravitational acceleration output from the simulation and the mass of the end clump, whereas the ram pressure force was estimated by $\Delta v^2 \rho S$ for a each cell with $S$ being the face area of the cell in this case, $\Delta v$ the velocity difference between neighbouring cells and $\rho$ the density in the given cell. The arrows show the deviation within 15 cells ahead of the core, the square represents the highest value and the end of the arrow the average. Although the determination of the ram pressure is uncertain, which is depicted by the large arrow, the graph supports the idea of a two phase collapse.
  
  With the equilibrium of forces a simple model can be established describing the collapse of a filament in two phases, given in the following section.

\section{Collapse of a uniform density filament}
\label{sec:collapse_uniform_density}

  Consider the simple case of a uniform density filament, where the acceleration along the main axis is well described by Equation \ref{eq:accelerationCollapseFilament}. A simple approach can be used to calculate the collapse of a filament.
  
  During the filament collapse the end core accumulates the swept up mass and, thus the mass increase of the core can be determined by
  \begin{align}
      \frac{\text{d}M}{\text{dt}} = \rho S v
  \end{align}
  with $M$ the mass of the core, $\rho$ the density, $S$ the face area of the core and $v$ its velocity (here we assume that the swept up gas is at rest). The core's equation of motion is governed by the gravitational self-acceleration at the filament's end which we denote as $a$:
  \begin{align}
      \frac{\text{d}(Mv)}{\text{dt}} = Ma
  \end{align} 
  using the product rule we get to
  \begin{align}
      v \frac{\text{d}M}{\text{dt}} + M \frac{\text{d}v}{\text{dt}} = Ma.
  \end{align}
  Using the mass increase defined above:
  \begin{align}
      M \frac{\text{d}v}{\text{dt}} + \rho S v^2 = Ma.
  \end{align}
  Already \citet{Clarke2015} saw that a terminal velocity is reached after a certain time and also our simulations show a linear collapse in the later stages. Thus, the collapse is divided into two parts: the accelerated part, where only the gravitational self-acceleration of the filament plays a role and the phase where the force equilibrium sets in and the core experiences a constant velocity. We assume the transition between the two phases to occur on a short time scale so that it can be described as a sharp phase transformation at turning time $t_{\mathrm{turn}}$ where $\text{d}v/\text{dt}\approx0$, which leads to:
  \begin{align}
      \rho S v^2 &\approx Ma
  \end{align}
  This results in a proportionality between the ram pressure force on the left hand side and the gravitational force on the right hand side 
  \begin{align}
      F_{\mathrm{ram}} &\propto F_{\mathrm{G}}
  \end{align}
  after the turning time. This seems reasonable, since Figure \ref{fig:length_force} shows the correlation between gravity and ram pressure after the turning time.
  
  Since the gravitational acceleration $a$ at the edge (Equation \ref{eq:accelerationCollapseFilament}) is independent of its length, a constant acceleration can be used. Thus, at the turning time $t_{\mathrm{turn}}$ the core has reached a velocity of $v=at_{\mathrm{turn}}$. Considering that the gas inside the filament is nearly at rest, while being swept up, this is exactly the velocity difference which produces the ram pressure after $t_\mathrm{turn}$. Assuming that the density ratio between the swept up material and the core is always similar at force equilibrium leads to
  \begin{alignat}{3}
      && (at_{\mathrm{turn}})^2S\bar{\rho} &\propto a S z \bar{\rho} \\
       \Rightarrow && t_{\mathrm{turn}} &\propto \sqrt{\frac{z}{a}}
  \end{alignat}
  with $z$ the size of the end region, which should be close to $2R$ assuming a symmetrical end region, and $S$ the face area of a slice of the filament. Using this and $\kappa$ as constant of proportionality, which is fitted to the simulation afterwards, results in:
  \begin{align}
      t_{turn} = \kappa \sqrt{\frac{2R}{a}}.
  \end{align}
  The acceleration is given by Equation \ref{eq:accelerationCollapseFilament}. Assuming that the end region is nearly symmetric the acceleration should be determined at $a\left(-\frac{l_0}{2}+\frac{z}{2}\right)$ ($l_0$ being the initial length of the filament), analogous to the determination of $t_{\mathrm{edge}}$ in \citet{Hoemann2021}:
  \begin{align}
      a\left(-\frac{l_0}{2}+\frac{z}{2}\right) = \alpha 2 \uppi G \bar{\rho} R
  \end{align}
  with $\alpha$ being a constant which also will be fitted to the simulation afterwards.
  
  Now, the time evolution of the length of the filament can be given by
  \begin{align} \label{eq:length}
      l(t) = \begin{cases}
      l_0-at^2 &\text{ for } t<t_{\mathrm{turn}}\\
      l_0+at_{\mathrm{turn}}^2-2at_{\mathrm{turn}}t &\text{ for } t\geq t_{\mathrm{turn}}
      \end{cases}
  \end{align}
  and the collapse time is defined as $l(t)=0$:
  \begin{align} \label{eq:Model}
      t_{\mathrm{col}} &= \frac{l_0}{2at_{\mathrm{turn}}}+\frac{1}{2}t_{\mathrm{turn}} \\
      &= \left( \frac{l_0}{2R\kappa} + \kappa \right) / \sqrt{\alpha 4 \uppi G \bar{\rho}}.
  \end{align}
  Using the aspect ratio $A$, given by $A=l_0/(2R)$, leads to
  \begin{align}
      t_{\mathrm{col}} = \left( \frac{1}{\kappa} A + \kappa \right) / \sqrt{\alpha 4 \uppi G \bar{\rho}}.
  \end{align}
  This is then the collapse time of a uniform density filament assuming the two phase model.
  
  In order to validate our approach and to fit $\kappa$ and $\alpha$ we performed hydrodynamic simulations as described in Section \ref{sec:simulating}. The influence of the different density profiles will be discussed in the next section, however, the collapse time agreed within 10\% between the simulations with uniform density and hydrostatic filaments. Since the hydrostatic profile is more realistic we used it in the following analysis. 
  Furthermore, we suppressed radial motions to prevent the filament to collapse radially because here we want to study the longitudinal collapse independently, influences are discussed in the next paragraph. The simulated filaments are defined by the aspect ratio $A$, the radius $R$ and the criticality $f$. The parameter space was covered as follows $f\in \{0.3, 0.5, 0.7\}$, $A \in \{8, 12, 15\}$, $R\in\{0.05\mathrm{\pc}, 0.10\mathrm{\pc}, 0.15\mathrm{\pc}\}$. We limited the simulations to those which were collapsed after a runtime of 4\Myr\ which is on the order of estimated filament lifetimes of a few Myr \citep{Andre2014}. Besides, we did simulations with $f\in\{0.1, 0.9\}$ but they sometimes show a different behavior than the other filaments, as for low line-mass filaments the edge cores accelerate to the centre faster then the real edge and thus a second pair of edge cores evolves and for large line-mass filaments material is swept up in the center which leads to an additional central core \citep[also seen in][]{Seifried2015}. 
  We thus had in total 17 simulations for evaluation, displayed in Figure \ref{fig:simulation}. The filament was considered as collapsed when the two density peaks merged and the aspect ratio of the filaments fell below 1. Each panel shows one of the parameters varied in the simulations on the x-axis and the collapse time on the y-axis. The second and third parameter are then plotted in different colours and styles, as indicated by the legends. The marker represent the simulated values. The lines show the fitted model. For clarity, not all simulations are plotted in one image. The model fits the simulated values very well for all configurations, with $\kappa = 1.22$ and $\alpha =0.69$. The collapse time is then given by:
  \begin{figure}
      \includegraphics[width=0.48\textwidth]{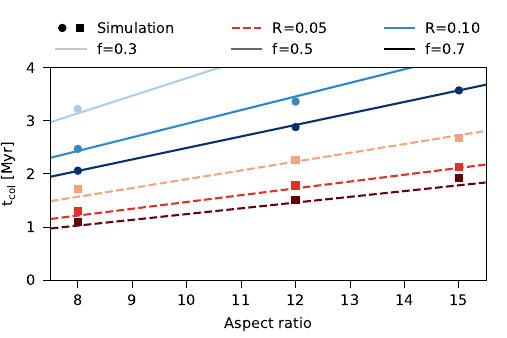}
      \includegraphics[width=0.48\textwidth]{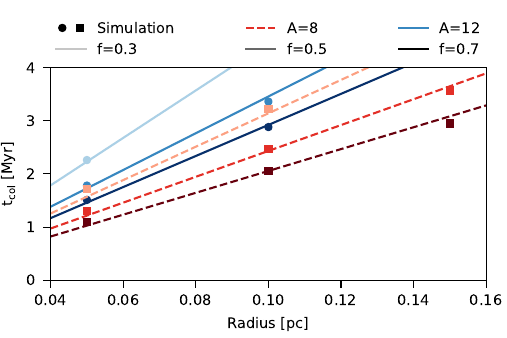}
      \includegraphics[width=0.48\textwidth]{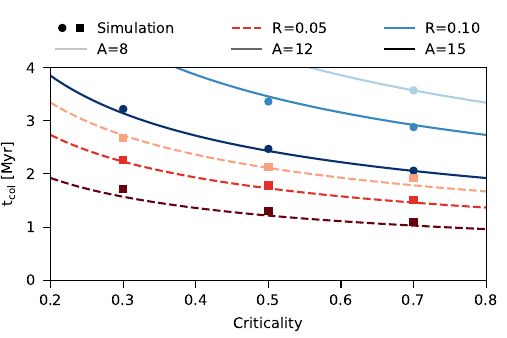}
      \caption{Results of the parameter study to determine $\kappa$ and $\alpha$ for Equation \ref{eq:Model}. Several simulations were carried out varying $f,R$ and $A$. The markers (dots and squares) show the collapse time of the simulation. The lines show the model for different parameters determined by colour and style. The model fits the simulations for $\kappa = 1.22$ and $\alpha =0.69$.}
      \label{fig:simulation}
  \end{figure}
  \begin{align}
      t_{\mathrm{col}} &= \frac{0.42+0.28 A}{\sqrt{G\bar{\rho}}} \\
      &= \frac{0.74 R+ 0.25 l_0}{\sqrt{G\mu}} \label{eq:collapse_mu}.
  \end{align}
  Besides small changes in the fitting factors this simple approach is also in agreement and explains Equation \ref{eq:clarke} by \citet{Clarke2015}. 
  
  Since we wanted to study the collapse along the main axis of the filament, the radial velocity was artificially set to zero to prevent the radial collapse. This was especially necessary for filaments having other profiles than the hydrostatic one. As these are not in hydrostatic equilibrium they would collapse or expand radially to adjust to the equilibrium profile. However, to show that the radial collapse itself has only minor influences on the collapse, we performed hydrodynamic simulations including also radial velocities perpendicular to the long axis. The results are depicted in Table \ref{tab:sinks}. As the cores start to collapse we included sink particles and define the time they need to reach the centre as the collapse timescale. As seen in Table \ref{tab:sinks}, the deviations between the simulations with and without radial velocity is at maximum one output and can be neglected. Thus, the suppression of radial velocities has no influence on the longitudinal collapse time of a filament. 
  
    \begin{table}
      \centering
      \caption{Comparison of collapse times determined in simulations with radial velocities $t_{\mathrm{col},v_r\neq 0}$ and without $t_{\mathrm{col},v_r=0}$. The time difference between the two simulations is given by $\Delta t$, whereas a time step in the simulation takes 0.07\Myr.}
      \label{tab:sinks}
      \begin{tabular}{cccccc}
        f   &   R [pc]   &   A   &   $t_{\mathrm{col},v_r=0}$ [Myr]  &   $t_{\mathrm{col},v_r\neq 0}$ [Myr] & $\Delta t$ [Myr] \\
        \hline
        0.3   &   0.05   &   12   &   2.26   &   2.26   &   0.00    \\
        0.5   &   0.05   &   12   &   1.78   &   1.71   &   0.07    \\
        0.7   &   0.05   &   12   &   1.51   &   1.44   &   0.07    \\
      \end{tabular}
  \end{table}

\begin{figure}
      \includegraphics[width=0.48\textwidth]{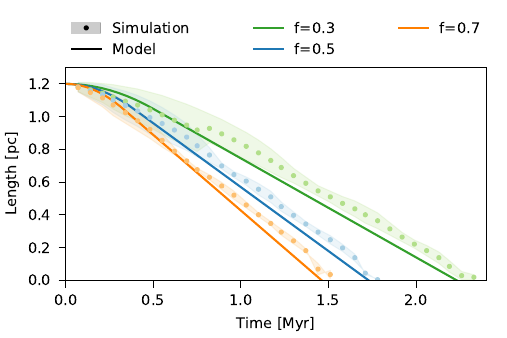}
      \caption{Comparison of the two phase model (solid lines, $A=12$ and $R=0.05$\pc) to the simulation values (light coloured as in Figure \ref{fig:length_force}). The filled area shows the region between the edge off the filament and its peak in the simulation, the mean is given by the dots. Different criticalities are displayed in different colours. The model reproduces the simulated collapses very well.}
      \label{fig:model_f}
  \end{figure}

  Not only does the model reproduce the collapse times accurately but it also describes the collapse itself as shown in Figure \ref{fig:model_f} where the length evolution of filaments with different criticalities are depicted. The shaded areas and the dots represent the simulation and the solid lines represent the corresponding model, as described in Figure \ref{fig:length_force}. In this example the aspect ratio was chosen to be $A=12$ and the radius to be $R=0.05$\pc. In all cases the simulation is very well reproduced by the model. The small underprediction is explained in the following section.
  
  All in all, the two phase model describes the collapse and the corresponding collapse time well and reproduces the empirical formula from \citet{Clarke2015}.
  
\section{Influence of different profiles}
\label{sec:adjustments}

  For the above analysis we assumed uniform density filaments since there is an analytical expression for the acceleration along a filament. Assuming a hydrostatic equilibrium, from a theoretical point of view we would expect to detect filaments with a hydrostatic profile which results in a $r^{-4}$ density profile in the outskirts. Although centrally peaked density profiles are observed in filaments their profile is flatter than the hydrostatic one, about $r^{-2}$ \citep{Lada1999, Arzoumanian2011, Palmeirim2013, Cox2016}. Thus, in the following we investigate the influence of different profiles on the collapse of a filament.

\begin{figure}
      \centering
      \includegraphics{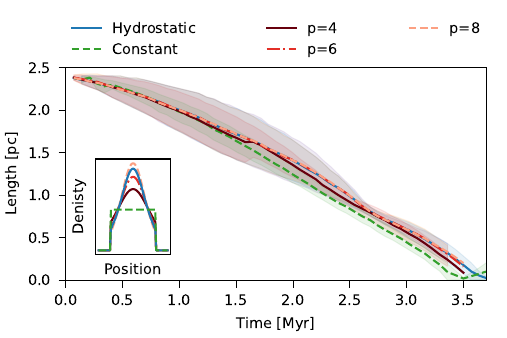}
      \caption{Length evolution for filaments with different radial density profiles with  $A=8, R=0.15$\pc\ and $f=0.5$. The hydrostatic profile is given in blue, Equation \ref{eq: Ostriker}, and steeper/flatter profiles according to Equation \ref{eq:profile} are given in shades of red with corresponding exponent p. In comparison the constant density profile is depicted in green. The shaded areas again show the difference between the core and the actual end of the filament. A comparison of the radial densities of the corresponding profiles is given in the subplot on the left hand side.}
      \label{fig:profile}
  \end{figure}

  Figure \ref{fig:acceleration} already shows that different radial profiles lead to different accelerations at the end of the filament. In order to check whether this has any kind of influence on the collapse time, we performed simulations with diverse density profiles. An example is given in Figure \ref{fig:profile}. All filaments have the same initial condition in terms of $A=8$, $R=0.15$\pc\ and $f=0.5$. The hydrostatic profile, given in blue, follows the density distribution for a filament in hydrostatic equilibrium (see Equation \ref{eq: Ostriker}). The label `Constant' means that the filament has a uniform density, represented by the green dashed line. The others follow a flatter or steeper profile than the hydrostatic one, a Plummer-like profile: 
  \begin{equation} \label{eq:profile}
      \rho(r) = \rho_\mathrm{c} \left[ 1 + \left( \frac{r}{H'} \right)^2 \right]^{-p/2},
  \end{equation}
  with $H'$ being:
  \begin{align}
      H'^2 = \frac{2c_s^2}{\uppi G \bar{\rho}}
  \end{align}
  and $\bar{\rho}$ the average density of the filament. The power-law index $p$ was varied in the range between four and eight. The central density is adjusted such, that the overall filament has a criticality of $f=0.5$. All filaments have a very similar time evolution (see Figure \ref{fig:profile}). However, a trend can be seen: Filaments with a flatter density distribution seem to collapse faster than filaments with a steeper density distribution. This is counterintuitive because the acceleration at the edge is larger for filaments with a more peaked profile (see Figure \ref{fig:acceleration}). We also see this trend in simulations with different parameters (in terms of $f,A$ and $R$). However, the effect seems to be always smaller than $10\%$. Thus, the approximation of the acceleration of a uniform density filament seems to be accurate enough.

  \begin{figure}
      \centering
      \includegraphics{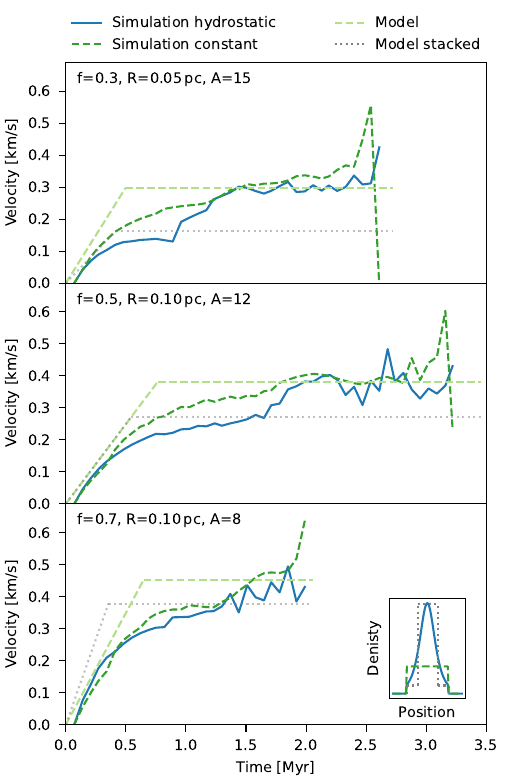}
      \caption{Velocity evolution for filaments with three different sets of parameters indicated in the left upper corner. Each graph compares the velocity evolution for the hydrostatic (blue solid line) and constant (green dashed line) velocity profile with the model (light green dashed line) and for a model with a stacked density distribution (grey dotted line). The profiles are depicted in the lower right corner.}
      \label{fig:velocity}
  \end{figure}

  In order to understand why the hydrostatic filament collapses slower compared to the filament with the constant density distribution, we look at their velocity evolution in the simulation given in Figure \ref{fig:velocity}. Three different examples are shown and their parameters are indicated in the upper left corner. The blue solid line indicates the velocity evolution of the hydrostatic filament, whereas the green dashed line depicts the uniform density case. Especially for lower line-mass filaments with f=0.3 and f=0.5 the velocity evolves differently for the two cases. Assuming the two phase model, a linear increase of velocity followed by a terminal velocity is expected as shown by the light green dashed line. It is a reasonable approximation of the simulated velocity for a constant density profile. However, for the hydrostatic case, there are strong deviations before reaching the final terminal velocity predicted by the model. An intermediate terminal velocity is reached. This can be explained by the strong density increase in the centre for a peaked profile, as can be seen in the box on the lower right side. Since the density inside is much higher, this induces a stronger ram pressure, and thus an earlier phase transition to an intermediate terminal velocity. We assume an approximation by a stacked profile would be more fitting, consisting of two constant density filaments, a more dense inner region and a diffuse outer region as indicated by the grey dotted line in the profile plot. The inner density is determined by the central density $\rho_c$, whereas the outer density is the boundary density $\rho_b$, the density cut is chosen such that the filament has the same criticality and radius as the other profiles. First the inner dense region collapses which reaches a terminal velocity rather fast but after a certain time the more diffuse outer region collapses onto the core and adds the momentum to get to the final terminal velocity. This effect delays the collapse for a short amount of time, and thus a systematic trend is seen that uniform density filaments collapse faster than hydrostatic ones. Because the deviations in the simulations were always below $10\%$, the effect is negligible.
  
  \begin{figure}
      \includegraphics[width=0.48\textwidth]{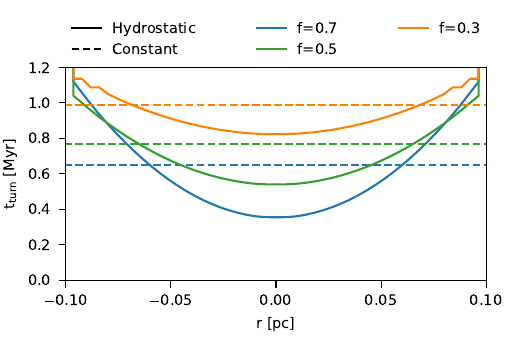}
      \caption{The radial dependency of the turning time for different criticalities. The dashed line represents a constant density distribution whereas the solid line shows the turning time of a hydrostatic profile. The used parameters are $A=8$ and $R=0.1$\pc.}
      \label{fig:turning_time}
  \end{figure}
  
  This effect is depicted for different criticalities in Figure \ref{fig:turning_time}. Since the turning time only depends on the density, it is constant for a uniform density filament (depicted by the dashed line), whereas the hydrostatic filaments show a radially dependent turning time. Although the constant profile gives a good median, it is only an approximation.
  
  Summing up, since the deviations for different filament profiles are small in simulations, it is sufficient to use the uniform density model in general as a good approximation of the collapse timescale.

\section{Discussion}
\label{sec:discussion}

  During this analysis several approximations have been made to describe the complex collapse of a filament. We will discuss in the following how these affect the presented results and give an example application of our model.
  
\begin{figure}
      \includegraphics[width=0.48\textwidth]{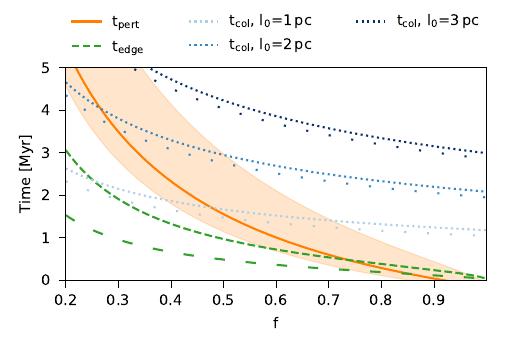}
      \caption{Comparison of different timescales in dependence of the criticality $f$. The edge effect formation timescale visualized in dashed green for $R=0.1$\pc\ (Equation \ref{eq:edge}, and the loosely dashed line for $R=0.05$\pc), perturbation timescale in orange (Equation \ref{eq:perturbation}, in the shaded area the perturbation strength is varied) and the collapse timescale in dotted blue (Equation \ref{eq:Model}). Varying initial filament lengths are given by different shades of blue. The dotted lines show the collapse time for $R=0.1$\pc\ and the sparsely dotted lines for R=$0.05$\pc.}
      \label{fig:petubation}
  \end{figure}

  During the overall collapse the edge effect is not the only way of creating cores inside the filament. Perturbations can grow inside also leading to collapsing regions. For a hydrostatic filament the timescale on which perturbations form cores inside the filament is given by the perturbation timescale 
  \begin{align}\label{eq:perturbation}
      t_{\mathrm{pert}} = \tau _{\mathrm{dom}} \log \left[ \left( \frac{1}{f} -1 \right) \frac{1}{\epsilon} \right],
  \end{align}
   which was already given by Equation 25 in \citet{Hoemann2021}. $\tau _{\mathrm{dom}}$ is the dominant fragmentation mode, which was calculated following Appendix E in \citet{Fischera2012}, compare \citet{Nagasawa1987, Heigl2020}. In the following we adopt a perturbation strength of $\epsilon=0.09$, based on the observations of \citet{Roy2015}. Figure \ref{fig:petubation} shows the perturbation timescale (orange solid line, the light orange region depicts $0.01<\epsilon<0.17$) in comparison to the edge effect formation timescale (Equation \ref{eq:edge}; dashed solid line for $R=0.1$\pc\ and loosely dashed for $R=0.05$\pc) and the collapse timescale for $R=0.1$\pc\ in dotted lines and $R=0.05$\pc\ in the loosely dotted lines for different lengths $l_0$. For low line-mass filaments with small aspect ratios $A$ a dominant edge effect can be expected, whereas for filaments with larger line-masses, perturbations grow faster. The edge effect, leading to dense cores on both ends of the filament, should however be visible in all cases.
  
  In this study we considered that the initial density distribution does not change along the filament. However, if we vary the density, different collapse modes can be expected. For example \citet{Seifried2015} found that a density enhancement of factor three leads to a `centralized collapse' mode.
  
  Since the presented model provides the length evolution of a filament, the age of a filament can be determined if the original length can be estimated. Taking into account the total mass of the filament and that the density in the inner part of the filament stays mostly constant, the length can be extrapolated. Using the current length, the age can then be determined by rearranging Equation \ref{eq:length}:
  \begin{align}
      t = \begin{cases}
      \sqrt{\frac{l_0-l(t)}{\alpha 2 \uppi G\bar{\rho}R}} & \text{for } t<t_{\mathrm{turn}}\\
      \left( \frac{l_0-l(t)}{\kappa 2R} + \kappa \right) / \sqrt{\alpha 4 \uppi G\bar{\rho}}& \text{for } t \geq t_{\mathrm{turn}}
      \end{cases},
  \end{align}
  considering that the edges are the only cores inside the filament, because the model does not account for other perturbations.
  
  Summing up, the presented model has limitations due to its approximations but it gives a good estimate of the collapse time and the length evolution, especially for low line-mass and short filaments.

\section{Conclusion}
\label{sec:conclusion}

  We show that the longitudinal collapse of a uniform density filament is a two step process. In the first phase, the filament accelerates until the turning time where the ram pressure counterbalances the gravitational acceleration and a terminal velocity is reached. This leads to a simple analytic model describing the length evolution and the resulting collapse time which explains the established empirical equation by \citet{Clarke2015}. For filaments with a peaked density distributions, the radial dependence of the turning time leads to the fact that the final terminal velocity is reached later. However, we find that all tested profiles are nevertheless well described by the collapse of a uniform density filament because the deviations are small. Since the collapse timescale is long in comparison to the edge effect formation and the perturbation timescale, it is unlikely that filaments collapses before forming cores. Thus, the edge effect is expected to dominate in low line-mass filaments whereas for large line-mass filaments perturbations can grow on a similar timescale.

\section*{Acknowledgements}

  We thank the anonymous referee for the detailed comments improving the quality of the paper. This research was supported by the Excellence Cluster ORIGINS which is funded by the Deutsche Forschungsgemeinschaft (DFG, German Research Foundation) under Germany’s Excellence Strategy - EXC-2094 - 390783311. We thank the CAST group for helpful discussion and comments.

\section*{Data Availability}

The data used for this analysis will be made available on request.
 



\bibliographystyle{mnras}
\bibliography{literature} 




\appendix




\bsp	
\label{lastpage}
\end{document}